# Fabricating Superconducting Interfaces between Artificially-Grown LaAlO$_3$ and SrTiO$_3$ Thin Films

*Danfeng Li,\* Stefano Gariglio, Claudia Cancellieri,† Alexandre Fête, Daniela Stornaiuolo*, and *Jean-Marc Triscone*

DPMC, University of Geneva

24 Quai Ernest Ansermet, 1211 Geneva, Switzerland

ABSTRACT: Realization of a fully metallic two-dimensional electron gas at the interface between artificially-grown LaAlO$_3$ and SrTiO$_3$ thin films has been an exciting challenge. Here we present for the first time the successful realization of a superconducting 2DEG at interfaces between artificially-grown LaAlO$_3$ and SrTiO$_3$ thin films. Our results highlight the importance of two factors—the growth temperature and the SrTiO$_3$ termination. We use local friction force microscopy and transport measurements to determine that in normal growth conditions the absence of a robust metallic state at low temperature in the artificially-grown LaAlO$_3$/SrTiO$_3$ interface is due to the nanoscale SrO segregation occurring on the SrTiO$_3$ film surface during the growth and the associated defects in the SrTiO$_3$ film. By adopting an extremely high SrTiO$_3$ growth temperature, we demonstrate a way to realize metallic, down to the lowest temperature, and superconducting 2DEG at interfaces between LaAlO$_3$ layers and artificially-grown SrTiO$_3$ thin films. This study paves the way to the realization of functional LaAlO$_3$/SrTiO$_3$ superlattices and/or artificial LaAlO$_3$/SrTiO$_3$ interfaces on other substrates.

Driven by the need for new building blocks for future electronic devices, intense effort is currently being devoted to the study of materials with new and/or multiple functionalities. Among them, transition metal-oxides, such as those with the perovskite structure, have been attracting a lot of attention due to the broad spectrum of their remarkable physical properties. Their structural compatibility allows them to be combined into complex heterostructures that serve as the ideal playground for exploring new phenomena resulting from the particular stacking or interface interactions.[1,2] One remarkable recent example is the observation of a two-dimensional electron gas (2DEG) at the interface between two band insulators, namely $LaAlO_3$ (LAO) and $SrTiO_3$ (STO).[3] Different features of this interfacial metallic state, such as its manifestation only for LAO layer thicknesses above or equal to 4 unit cells (u.c.),[4] can be elegantly explained by the so-called *polar-catastrophe* model: the polar discontinuity between the polar LAO and the non-polar STO induces a transfer of electronic charges from the surface to the interface.[5–8] The physical properties observed in this two-dimensional electron system have generated a lot of interest and widespread research activity.[9–12] Recently, field-effect transistors[13] and nano-devices[14,15] have been realized based on this system, show-casing its potential for nano-electronic applications.

One major step towards possible applications in novel devices is the integration of this 2DEG with other functional materials, like widely used semiconductors. Moreover, transferring this interface onto other single-crystalline perovskite substrates, that possess higher crystalline quality, is an exciting challenge since it could potentially result in interfaces with higher carrier mobility.[16] This approach also opens possibilities for exploiting the strain effect, which in the LAO/STO interface still awaits full exploration.[17] More interestingly, the capability to build the LAO/STO interface on artificially-grown STO is also fundamental for realizing LAO/STO superlattices. In fact, a few reports have discussed different physical aspects of LAO/STO superlattices,[18,19] whereas a clear experimental demonstration of multiple parallel conducting 2DEGs is still elusive. Such heterostructures would allow, for instance, the study of the coupling between 2D superconductors.[20] In fact, the major challenge to achieve these goals is to obtain a high quality metallic interface on an artificially-grown STO film. To date, metallic two-dimensional interfaces down to low temperatures have only been observed when $LaAlO_3$ is grown directly on single-crystalline $SrTiO_3$ substrates. Efforts have been made towards realizing such interfaces between LAO and artificially-grown STO films and conducting LAO/STO interfaces grown on other substrates have been achieved.[21–24] However, the metallic behavior was no longer maintained in the low temperature range and localization effects were observed.[23] The main obstacle remains to be preserving both the metallicity down to the lowest temperature and the two-dimensional nature of the electron gases by using proper oxygen deposition conditions.[23]

In this article we explore the fabrication of an artificial LAO/STO interface, where the 2DEG is formed within a STO layer that was homoepitaxially grown on a (001) STO single-crystal. We demonstrate that the 2DEG at the fabricated interfaces is not only metallic down to the lowest temperatures, but also superconducting. Analysis of the transport properties shows that the growth temperature determines the crystalline perfection of the STO layer and consequently the metallicity of the system. Using Friction Force Microscopy (FFM), X-ray diffraction and transport measurements, we relate the absence of metallic behavior normally observed in LAO/STO thin films to the locally nanoscale SrO segregation on the STO surface and to the associated defects in the STO thin films. Our findings show that by controlling the top surface termination of the STO layer and optimizing the STO growth, the 2DEG can be preserved, offering a promising approach to realizing functional LAO/STO multilayers and/or superconducting LAO/STO interfaces on other substrates.

LAO and STO films were grown by pulsed-laser deposition (PLD) from single-crystalline targets. The laser fluence was set to ~0.6 J cm$^{-2}$ and the repetition rate was kept at 1 Hz. The growth was monitored by reflection high-energy electron diffraction (RHEED). We followed the evolution of the intensity of the specular spot to identify the growth mode. The STO films were grown on TiO$_2$-terminated (001)-oriented STO single-crystalline substrates at 800°C in an oxygen pressure of $8\times10^{-5}$ Torr and at 1100°C in an oxygen pressure of $1\times10^{-6}$ Torr. The LAO layers were grown at 800°C in an oxygen pressure of $8\times10^{-5}$ Torr. The samples characterized in this work are classified into two categories, depending on the fabrication process. For convenience, we name them *in-situ* samples and *ex-situ* samples. As shown in Figure 1, for *in-situ* samples, a STO thin film is grown on a (001) TiO$_2$-terminated STO substrate, followed by the deposition of a LAO film without breaking the vacuum. For the *ex-situ* samples, the following fabrication steps were employed: a) STO film deposition on a (001) TiO$_2$-terminated STO substrate; b) *ex-situ* surface treatment of the STO layer (buffered-HF etching followed by a high-temperature annealing procedure);[25,26] c) LAO film deposition. Extra care was taken to avoid as much as possible the formation of oxygen vacancies, both in the STO films and in the substrates, by performing an oxygen-annealing step during sample cooling in the deposition chamber. In this manner, we fabricated *in-situ* and *ex-situ* samples with STO thin films of different thicknesses (from 2 u.c. to 40 u.c.) grown at different temperatures (800°C and 1100°C), while keeping the LAO thickness constant (5 u.c.). All the scanning probe measurements were performed with a *Cypher* microscope (Asylum Research) in contact mode. Sb-doped Si cantilevers with a nominal force constant of k = 0.2 N m$^{-1}$ were used. The friction contrast was then obtained by a subtraction of the lateral retrace signal from the lateral trace. No attempts have been made to quantify the absolute friction force since the lateral force constant of the cantilever is unknown. The magneto-transport measurements were performed in a He$^4$ cryostat equipped with a superconducting magnet (8 T). Field-effect devices were prepared by depositing a gold pad as the gate electrode on the backside of the STO substrate. The superconducting transitions were measured in a He$^3$/He$^4$ dilution refrigerator with a base temperature of 30 mK.

We first focus on the *in-situ* and *ex-situ* samples with STO layers grown at 800°C. At this substrate temperature, STO grows layer by layer, as revealed by periodic RHEED intensity oscillations.[27] Figure 2a displays a sketch of a sample patterned into Hall-bar geometry for transport measurements. Figure 2b shows the room-temperature sheet conductance as a function of the STO layer thickness (*d*, in u.c.), for both the *in-situ* and *ex-situ* samples. From the figure, one can clearly see that for *in-situ* samples above *d* = 8 u.c. the system begins to lose conductivity: the sheet conductance falls outside the range of standard LAO/STO interfaces indicated by the dashed area, and drops below the measurement limit for *d* = 15 u.c. For *ex-situ* samples, the *d* = 15 u.c. sample is still metallic and thus the critical STO thickness for metallicity shifts upwards. This shift is also reflected in the sheet carrier density estimated from Hall effect measurements. Figure 2c shows -1/e$R_H$ ($R_H$ being the Hall constant) for the same series of samples. We note that, for the *ex-situ* samples, there is a clear tendency towards a carrier density reduction before they become immeasurable as the STO layer thickness is increased.[27] The comparison between *in-situ* and *ex-situ* samples highlights the importance of this surface treatment, suggesting that the $TiO_2$ termination may not be preserved during the growth of the STO films at 800°C.[28] In this case, regions with SrO termination could reduce the global conductivity of the system as evidenced by previous work where the deposition of a SrO layer on the STO surface was shown to inhibit the formation of the 2DEG at the interface.[5,26,29]

To investigate the possible presence of SrO on the surface, we used the friction force mode (FFM) of an atomic force microscope. FFM probes the nanoscale friction force as the tip is dragged across the scanned surface by monitoring the cantilever torsion during the scan. By subtracting the retrace scan from the trace scan of the lateral signal, one can disentangle the contribution due to the different friction coefficients from the one purely related to the topography, such as unit-cell step edges, thus obtaining a clear contrast with nanoscale resolution linked to the local terminations.[27] This approach has been previously employed to determine the local termination of complex-oxide surfaces and in particular to probe the partial SrO coverage on the surface of STO single-crystals.[30,31]

Figures 3a and 3b show the surface topography and the FFM images, respectively, of a single-crystalline (001) STO substrate. Unit-cell-high step-and-terrace structure, a characteristic feature of a miscut $TiO_2$-terminated STO surface, can be clearly seen on the topographic image. Despite these unit-cell steps, no contrast is visible on the friction image, suggesting uniform $TiO_2$ coverage. On top of such substrates, we grew 15 u.c. thick STO films, since for this layer thickness the *in-situ* and *ex-situ* LAO/STO/STO structures showed fundamentally different transport properties (insulating and metallic respectively). After the STO deposition, the surface becomes rougher, marked by the presence of nanometer-sized pits, islands and rough step edges, as shown in Figures 3c and 3e. This may already indicate that the $TiO_2$ termination is no longer well maintained and mixed termination appears. However, this feature is not resolved in the friction images shown in Figures 3d and 3f, where no clear contrast is distinguishable. This may be due to the fact that the termination of the surface is mixed on sub-nanometer scale that is below the resolution of FFM.

In order to detect the presence of SrO, we performed an annealing treatment in an oxygen atmosphere at ambient pressure on one 15 u.c. thick STO film. Such an annealing promotes the migration of surface species to the step edges,[32,33] and is routinely used as a complementary step after the buffered HF (BHF) etching to obtain the atomically flat $TiO_2$ termination with sharp unit-cell step edges. Since a long high-temperature anneal may also promote SrO segregation on the STO surface,[34] we keep our annealing time below 2 hours. Figures 3g and 3h illustrate the topographic and friction images after the annealing treatment. In the friction scan, we observe the appearance of a clear contrast, with nanometer-sized white areas present at the terrace edges that we relate to the accumulation of SrO. In order to exclude the possible segregation of SrO during the annealing step, we performed a control experiment by treating an identical 15 u.c. thick STO film first with BHF and then with the same annealing process. The topography and friction images illustrated in Figures 3i and 3j do not suggest any presence of SrO as we do not observe any contrast in the FFM scan. Therefore using these treatment conditions, once the segregated SrO is removed by the BHF etching, the $TiO_2$ termination is preserved during the high-temperature annealing.

We can then conclude that the SrO segregation occurs on the film surface during STO growth and may be responsible for the loss of the 2DEG at the interface for *in-situ* samples. We should also note that this loss of the 2DEG properly highlights the fact that oxygen vacancies, which can possibly be formed during the growth,[23] seem not to be the relevant origin of the metallicity in this study. Looking at Figs. 2b and 2c, however, raises the question: why is this loss also observed for *ex-situ* samples, albeit at higher STO critical thickness?

To address this question, we have analyzed the structural quality of the STO films grown at 800°C. Figure 4a shows a θ-2θ X-ray diffractogram around the (002) Bragg reflection of STO. Around the substrate peak we observe oscillations attributed to the finite thickness of the layer. A fit to these oscillations yields a film thickness of 40 u.c., in agreement with the estimation from the count of the RHEED oscillations.[35] The presence of these oscillations suggests a slight off-stoichiometry of the STO film,[36] probably originating from the SrO migration and segregation to the surface.[37] These defects in the STO films could induce localization of the 2DEG, resulting in an insulating interface. In order to improve the quality of the STO films, we have raised the growth temperature to 1100°C. Previous work has shown that above 1000°C, STO films grow by step-flow and have dielectric properties comparable to those in single crystals.[38] Indeed, during the deposition, we observe that the RHEED specular intensity recovers completely after each laser pulse, indicating step-flow growth.[27] X-ray diffraction data for samples grown at high temperature reveal only the substrate peaks with no finite size oscillations. In Figure 4a we compare a θ-2θ scan around the (002) reflection for the 800°C sample and for the 1100°C sample, both being 40 u.c. thick. The absence of the finite-size oscillations indicates that the film is indistinguishable from the single crystalline substrate. The AFM investigation of the 40 u.c. film surface before and after a thermal treatment (but without BHF etching) is summarized in Figures 4 c,d and e,f, respectively. The topography images shows atomically flat surface while no friction contrast is observed in the FFM images, indicating the absence of SrO segregation for these films.

We prepared a series of *ex-situ* samples with STO layers of different thickness grown at 1100°C. In Figure 4b, the room-temperature sheet conductance of these samples is compared with that of *ex-situ* samples grown at 800°C. One can see that for STO layers grown at 1100°C, the 2DEG is preserved for STO films with a thickness of 40 u.c. or about 16 nm (the maximum thickness we used). This value exceeds the spatial extent of the 2DEG, which has been shown to be 1-2 u.c. at room temperature[39,40] and a few nanometers at low temperature.[39,41] All the samples with STO layers grown at 1100°C show metallic behavior down to the lowest temperature (1.5K). Finally, we also fabricated one *in-situ* sample with a STO thickness of 40 u.c., which also displays the fully metallic behavior and whose room-temperature sheet conductance is also indicated in Figure 4b. This suggests that increasing the STO growth temperature not only preserves the STO crystal quality but also its $TiO_2$ surface termination.

Figure 5 shows a detailed characterization of the transport properties of *ex-situ* LAO/STO samples with a 40 u.c. thick STO layer grown at 1100°C. The values of the carrier density (estimated via the Hall effect) and of the electron mobility shown in Figure 5a,b are in line with standard LAO/STO interfaces.[42] In particular, we do not observe a suppression of the carrier mobility and a strong localization behavior, as reported in Ref.[23]. However, we also do not observe an enhancement of the carrier mobility. In order to further characterize this artificial interface, field-effect experiments in back-gate geometry were performed; in this configuration, the STO layer and the STO substrate act as the series gate dielectrics. The large tunability of the sheet resistance and the evolution of the magneto-resistance with gate voltage shown in Figure 5c are very similar to those of standard LAO/STO interfaces. These results confirm that high temperature growth enables the fabrication of high quality interfaces. Superconductivity was also observed at these interfaces made of artificially-grown LAO and STO thin films in the millikelvin temperature range. Figure 5d illustrates the normalized resistance versus temperature for a standard LAO/STO interface and for a sample with a 40 u.c. thick STO layer grown at 1100°C. A clear and complete superconducting transition is observed for the STO thin film sample albeit with a lower $T_c$ of about 100mK. This can be related to a different doping state and may suggest differences in the superconducting state that need to be fully explored. Given that the thickness of the superconducting electron gas at LAO/STO interfaces is ~10 nm,[41] these results demonstrate, for the first time, the successful generation of a two-dimensional superconducting electron gas between artificially-grown LAO and STO thin films. These results open a new pathway to realize the LAO/STO superlattices with multiple parallel 2D superconductors.

In conclusion, we have presented a successful realization of a superconducting 2DEG at interfaces between artificially-grown LAO and STO thin films. Our results highlight the importance of two factors—the growth temperature and the STO termination. Samples grown at 800°C do not host the 2DEG for STO layer thicknesses above a critical value. The observation of nanoscale SrO segregation on the STO surface by the local friction force microscopy and the presence of defects in the STO layer seems to be plausible origins for this behavior. Increasing the deposition temperature for the STO layers to 1100°C changes the growth mode to step-flow and produces films of higher quality; a metallic, down to the lowest temperatures, and superconducting 2DEG is observed on these artificial LAO/STO interfaces even for thick STO layers. This study provides the key ingredients needed to grow LAO/STO superlattices or interfaces on other substrates, where exciting physical properties can be realized and studied.

**Corresponding Author**

*E-mail: Danfeng.Li@unige.ch

**Present Addresses**

†Paul Scherrer Institut, CH-5232 Villigen, Switzerland

ACKNOWLEDGMENT

We are deeply grateful for the fruitful discussions with Mathilde L. Reinle-Schmitt and to Benedikt Ziegler for the assistance with FFM measurements. We also thank Marco Lopes and Sébastien Muller for the technical assistance. This work was supported by the Swiss National Science Foundation through the National Center of Competence in Research, Materials with Novel Electronic Properties, MaNEP and division II.

**Figure 1.** Schematic of the fabrication processes for the so-called *in-situ* and *ex-situ* samples.

**Figure 2.** Transport properties of the LAO/STO/STO samples with STO layers grown at 800°C. a) A sketch of the LAO/STO/STO sample with a Hall-bar pattern used to measure the sheet resistance and carrier density of the system. b) Sheet conductance at 280 K; c) -1/e$R_H$ versus STO layer thickness $d$ (u.c.) at 5 K. Black open (blue solid) circles are for *in-situ* (*ex-situ*) samples. The shaded areas in b) and c) indicate the room-temperature sheet conductance and -1/e$R_H$ at 5 K respectively in standard LAO/STO interfaces. The lines are guides to the eye.

**Figure 3.** Surface characterization a), b) of a TiO$_2$-terminated STO substrate, and c)-j) of 15 u.c. thick STO films grown at 800°C using atomic force microscopy. Vertical force microscopy (VFM) images display contrast due to surface topography, whereas friction force microscopy (FFM) images exhibit chemical contrast due to the mixed termination of the surface. c), e) VFM and d), f) FFM images of two as-grown 15 u.c. STO films. g) VFM and h) FFM images of the STO film surface after the annealing at 1000°C. The contrast in the FFM image reveals the presence of a mixed-terminated surface. i) VFM and j) FFM images of a 15 u.c. STO film etched by BHF and annealed at 1000°C. No contrast in the FFM scan is observed.

**Figure 4.** a) X-ray diffraction patterns of 40 u.c. homoepitaxial STO films grown at 800°C and 1100°C. The arrows indicate finite size oscillations. b) Sheet conductance at 280 K for STO layers of different thickness $d$ (u.c.). Blue circles (black squares) are for *ex-situ* samples with STO layers grown at 800°C (1100°C). The red star is for an *in-situ* sample with the STO layer grown at 1100°C. The shaded area indicates the room-temperature sheet conductance in standard LAO/STO interfaces. The lines are guides to the eye. c) VFM and d) FFM images of a 40 u.c.

thick STO film grown on a TiO$_2$-terminated STO substrate at 1100°C. e) VFM and f) FFM images of the same sample after annealing at 1000°C.

**Figure 5.** Magnetotransport properties and superconductivity of a LAO/STO/STO heterostructure with a STO layer of 40 u.c. grown at 1100°C. a) Inverse Hall constant and b) the electron mobility as a function of temperature. c) Magnetoresistance at 1.5 K for different gate voltages V$_g$. d) Normalized resistance *R*(T)/*R*(500mK) as a function of temperature displaying superconducting transitions. Blue solid squares and black open circles are for a LAO/STO interface with artificially-grown STO layer and for a "standard" LAO/STO sample respectively.

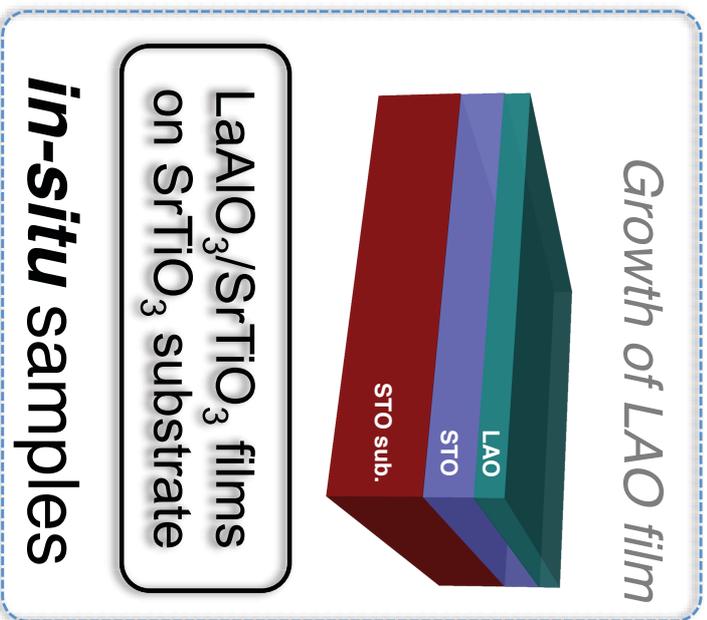
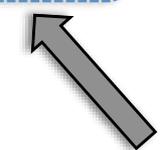
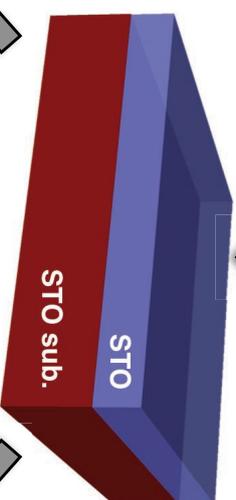
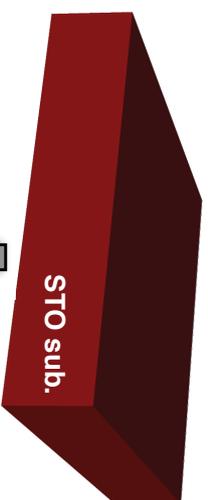
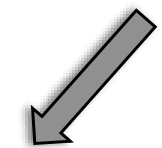
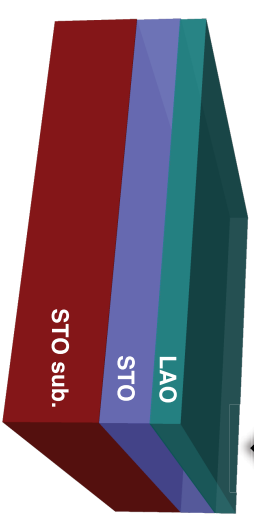

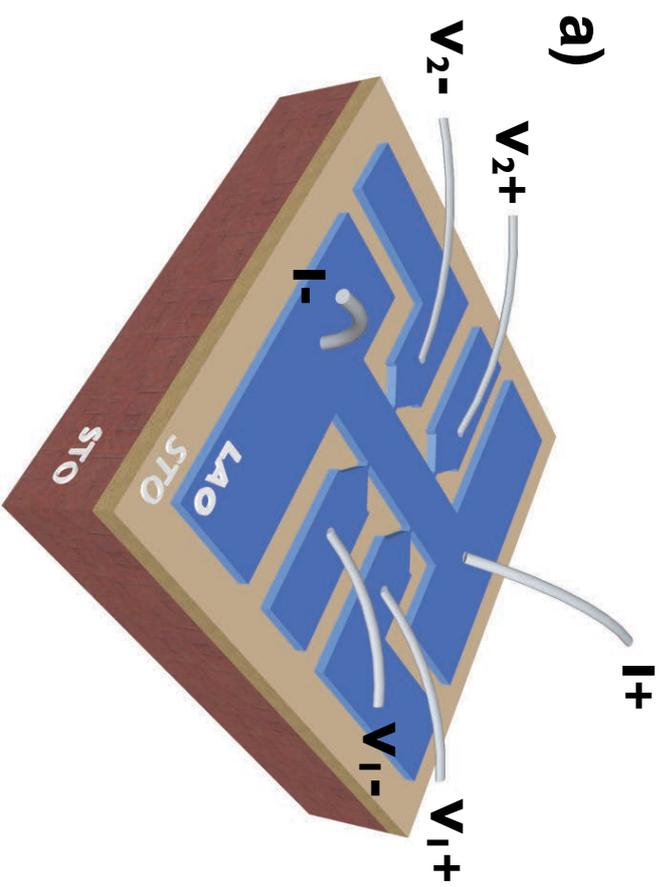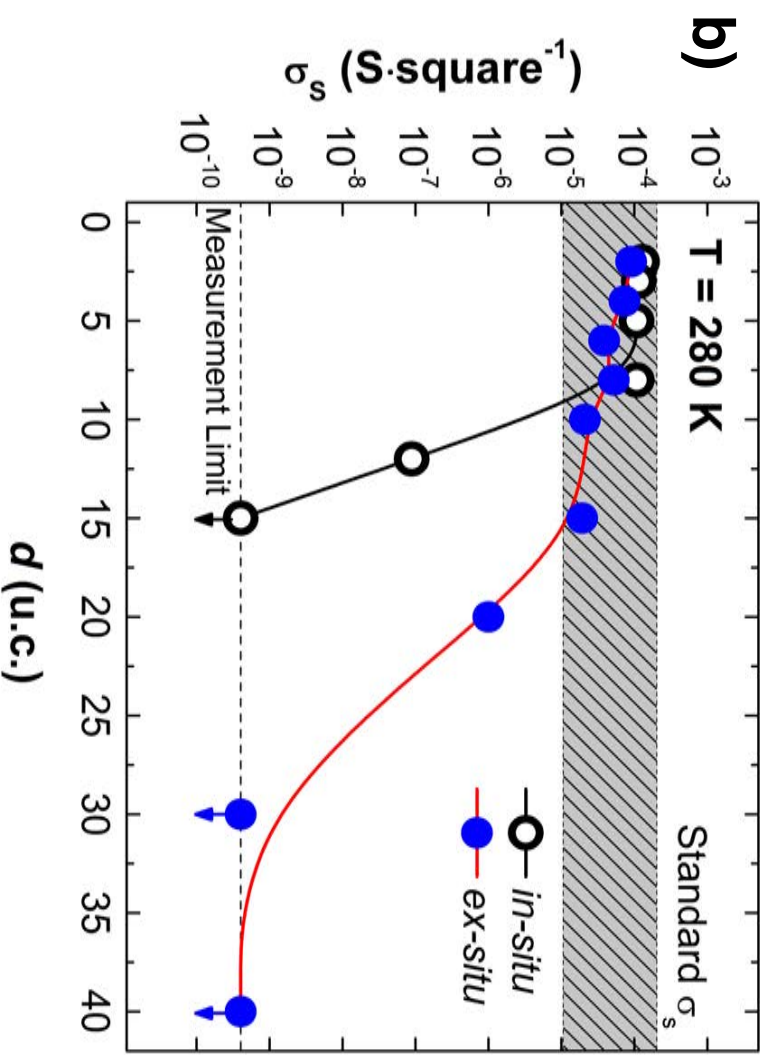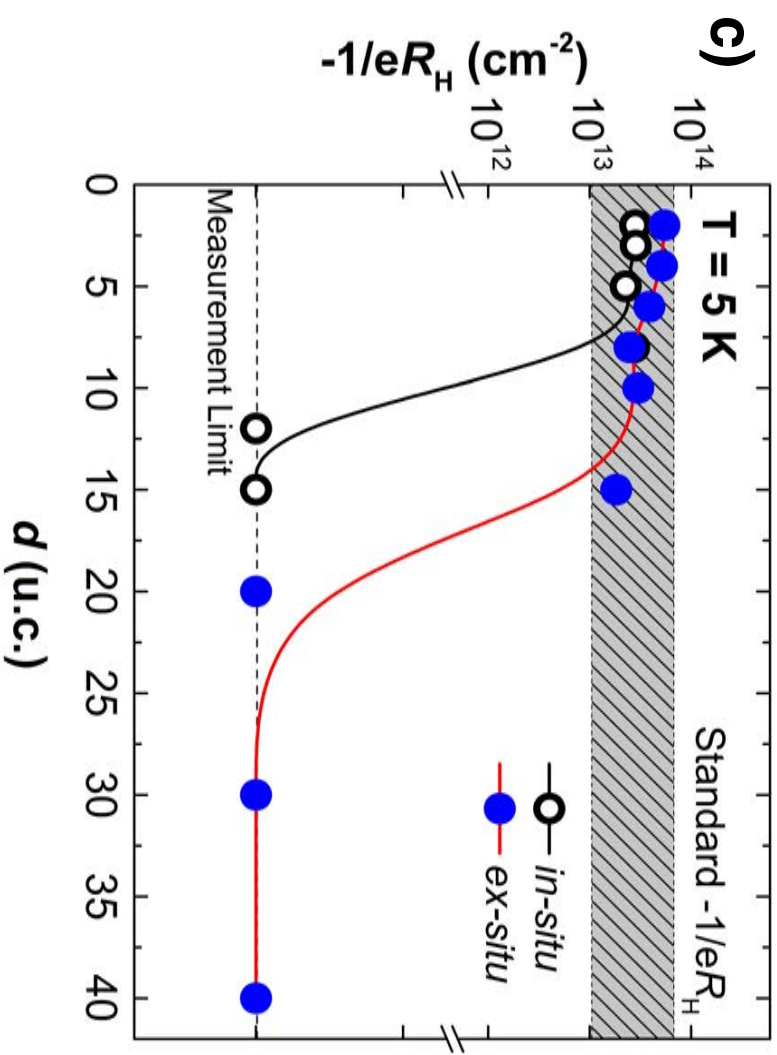

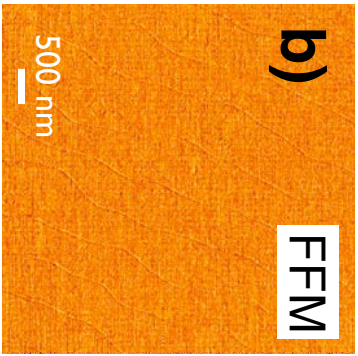
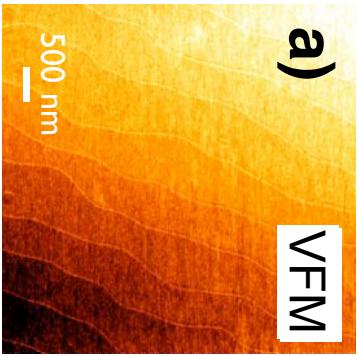
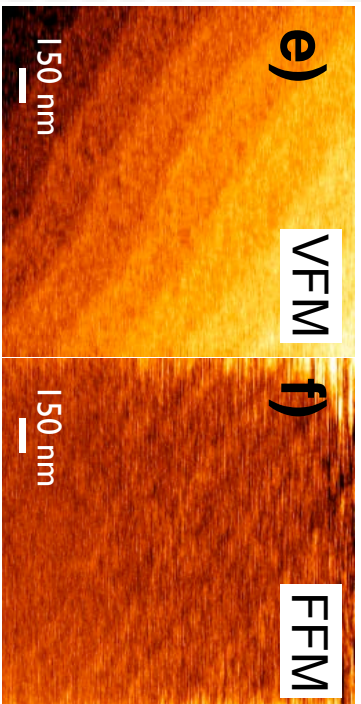
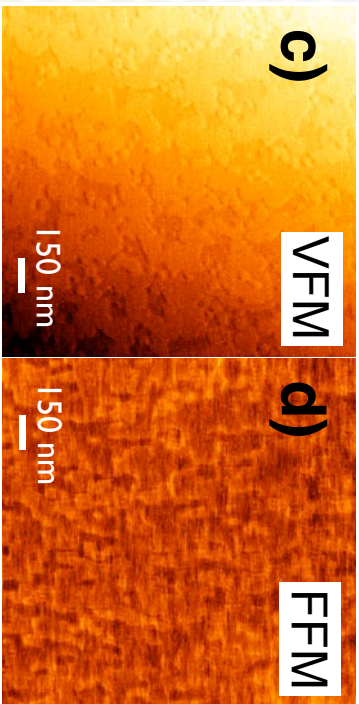
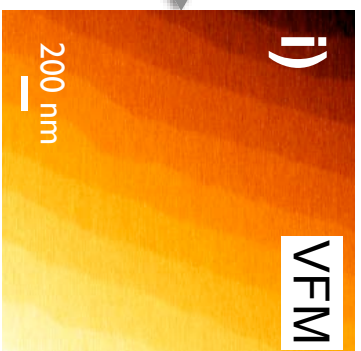
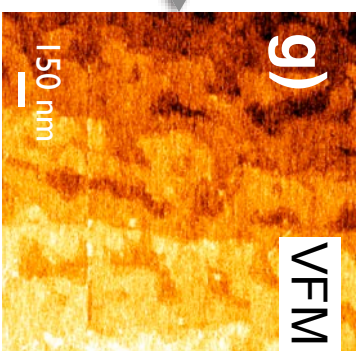
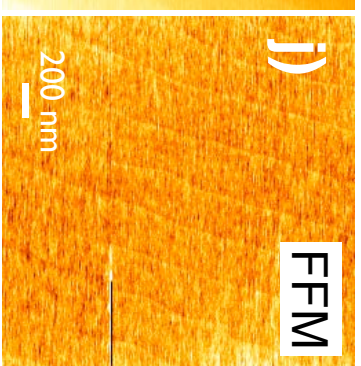
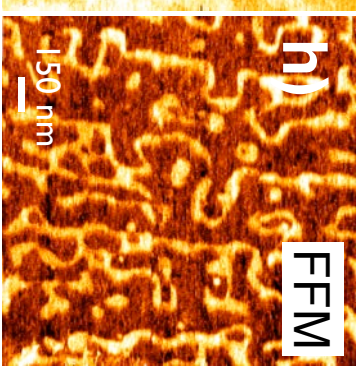

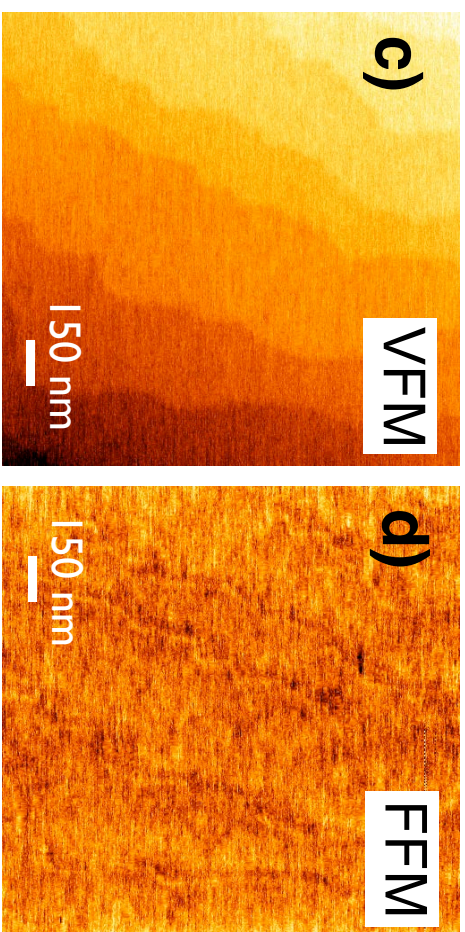
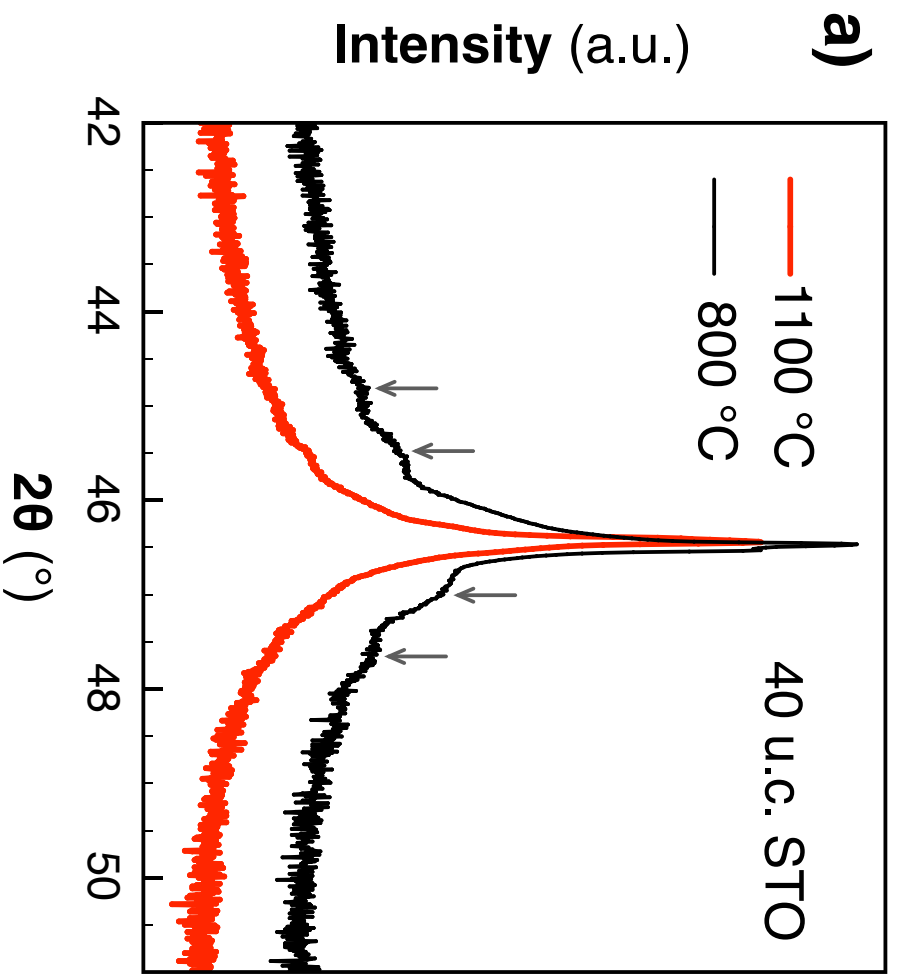
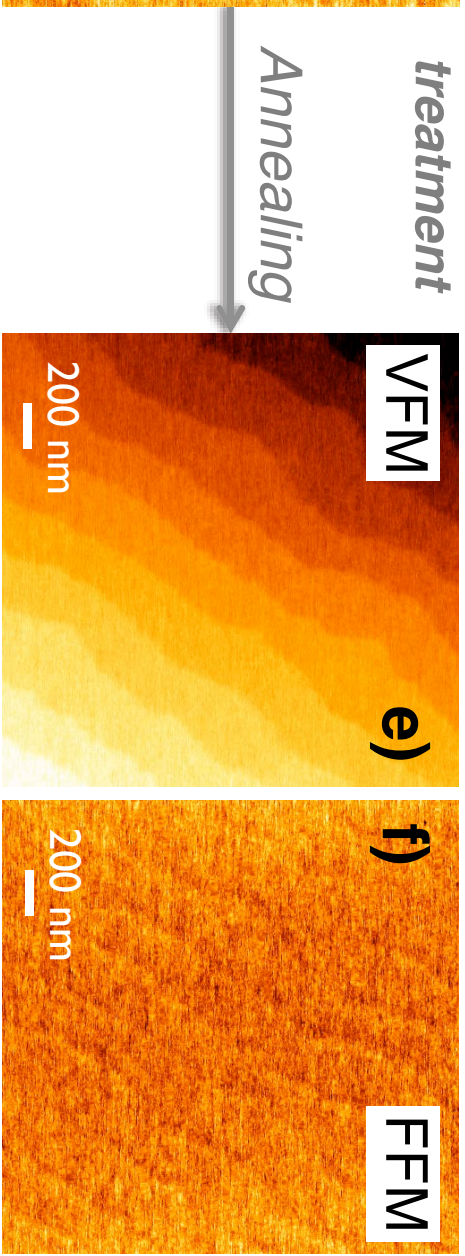
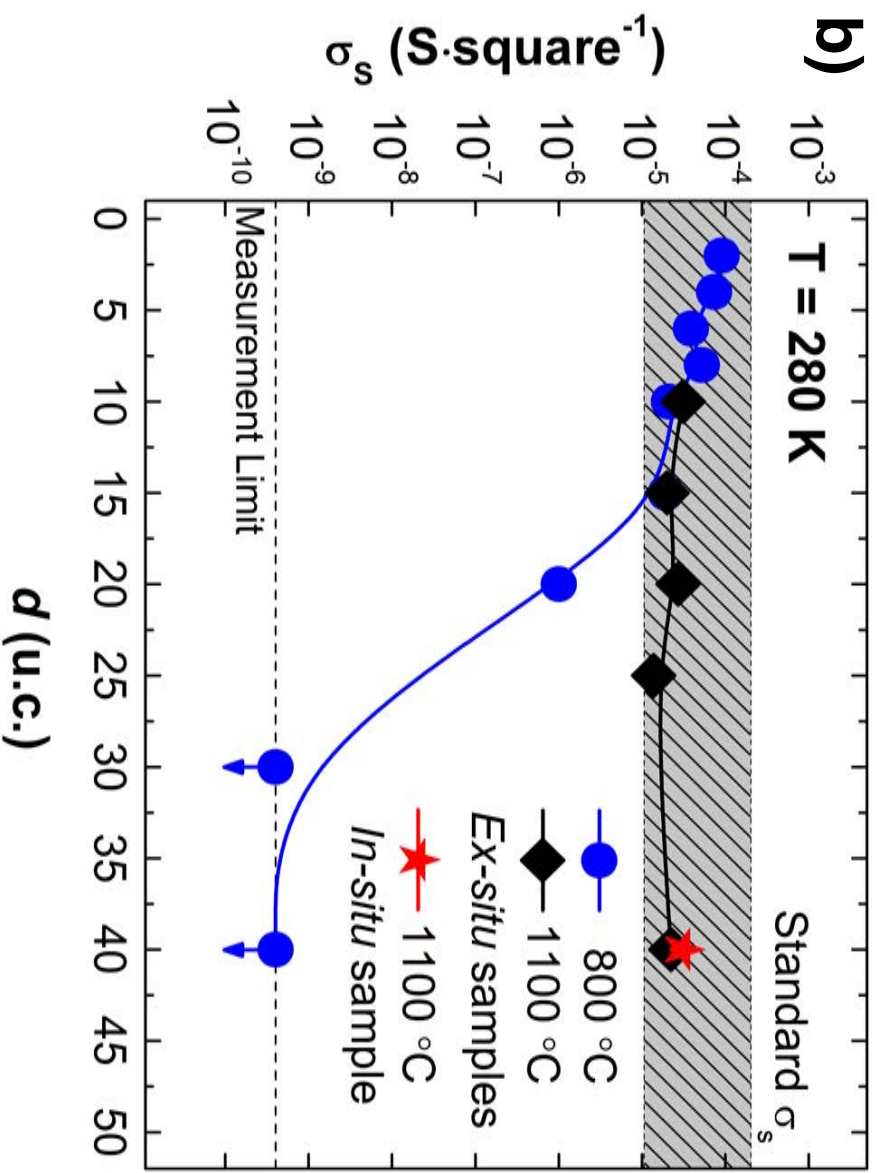

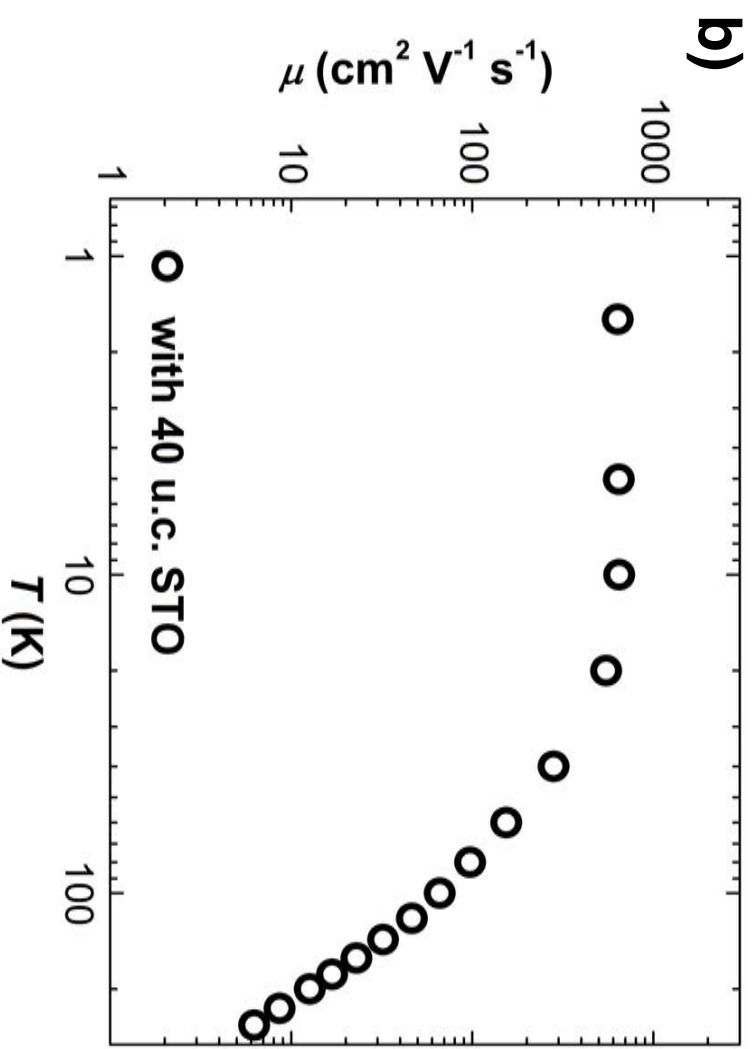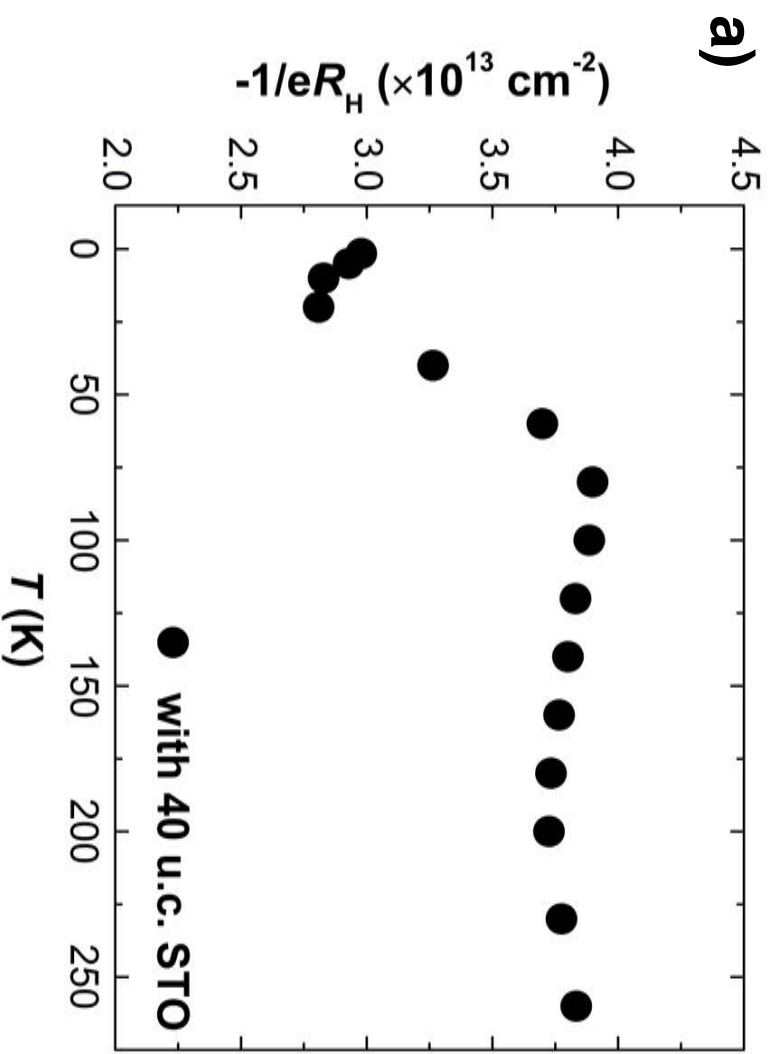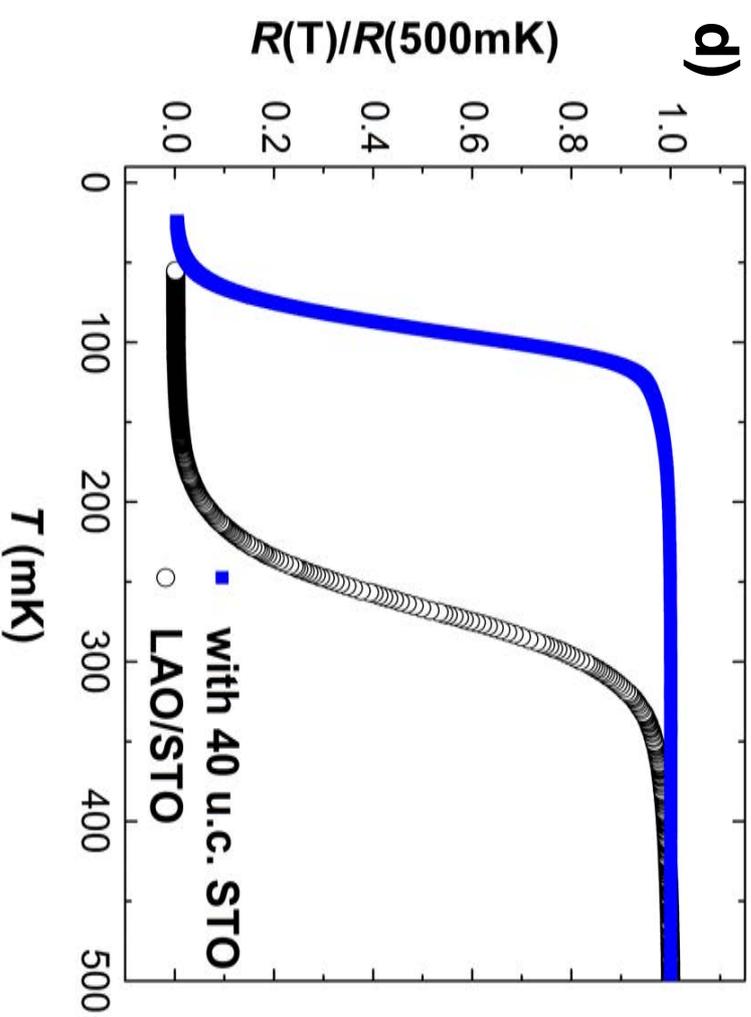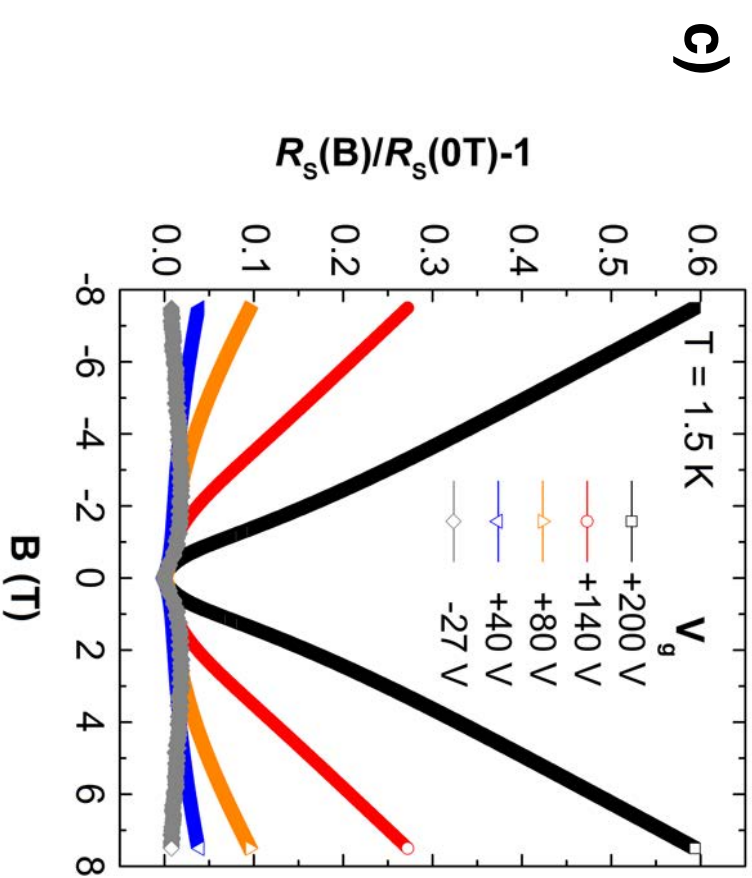

# Supporting Information

Fabricating Superconducting Interfaces between LaAlO$_3$ and Artificially-Grown SrTiO$_3$ Thin Films

*Danfeng Li,\* Stefano Gariglio, Claudia Cancellieri,† Alexandre Fête, Daniela Stornaiuolo, and Jean-Marc Triscone*

**Experimental**

*Sample growth*: LAO and STO films were grown by pulsed-laser deposition (PLD) from single-crystalline targets. The laser fluence was set to ~0.6 J cm$^{-2}$ and the repetition rate was kept at 1 Hz. The growth was monitored by reflection high-energy electron diffraction (RHEED). We followed the evolution of the intensity of the specular spot to identify the growth mode. The STO films were grown on TiO$_2$-terminated (001)-oriented STO single-crystalline substrates (from Crystec GmbH) at 800°C in an oxygen pressure of 8×10$^{-5}$ Torr and at 1100°C in an oxygen pressure of 1×10$^{-6}$ Torr. The LAO layers were grown at 800°C in an oxygen pressure of 8×10$^{-5}$ Torr. For *in-situ* samples, the LAO layer was grown immediately after the STO film; for *ex-situ* samples, the LAO layer was grown after the surface treatment to regain the TiO$_2$ termination. At the end of every deposition, the sample was annealed *in-situ* at 500°C in an oxygen pressure of 200 mbar for one hour and cooled down in the same atmosphere.

*Surface treatment*: Commercial buffered hydrofluoric acid (BHF) (from Merck) was used as the chemical etchant to remove the SrO from the STO surface [17]. Before etching, the samples were cleaned in de-ionized water in an ultrasonic bath. The samples were then immersed in the BHF solution for 30-60 seconds, followed by a de-ionized water and ethanol cleaning. An annealing treatment was performed in a tube furnace with flowing O$_2$ at 1000°C for 1-2 hours in order to promote surface migration [28,29]. Heating and cooling rates of 10°C/min and 5°C/min were used, respectively.

*Friction force microscopy*: All the scanning probe measurements were performed with a *Cypher* microscope (Asylum Research) in contact mode. Sb-doped Si cantilevers with a nominal force constant of k = 0.2 N m$^{-1}$ were used. The contact force was kept at about 50 nN. The lateral deflection was simultaneously monitored by the four-quadrant photo-detector. The friction contrast was then obtained by a subtraction of the lateral retrace signal from the lateral trace. No attempts have been made to quantify the absolute friction force since the lateral force constant of the cantilever is unknown. In a friction force image, bright colors show high friction forces while dark colors indicate low friction forces.

*Transport characterization*: The electrical transport measurements were performed in a $He^4$ cryostat equipped with a superconducting magnet (8 T). Field-effect devices were prepared by depositing a gold pad as the gate electrode on the backside of the STO substrate. The superconducting transition was measured in a $He^3/He^4$ dilution refrigerator with a base temperature of 30 mK.

**SrTiO$_3$ growth and structural characterization**

The film growth was monitored *in-situ* using reflection high-energy electron diffraction (RHEED) by following the intensity oscillations of the specular spot. After the SrTiO$_3$ (STO) growth, topography images were obtained using atomic force microscopy in order to check the film surface quality. Figure S1 shows the RHEED intensity oscillations and the topography of the 10 u.c., 20 u.c. and 30-u.c.-thick STO films grown at 800°C on TiO$_2$-terminated STO substrates: The RHEED intensity oscillations evidence layer-by-layer growth. Figure S2 shows the RHEED intensity and the topography of a 40-u.c.-thick STO film grown at 1100°C on TiO$_2$-terminated STO substrates: The growth mode is step-flow at this temperature.

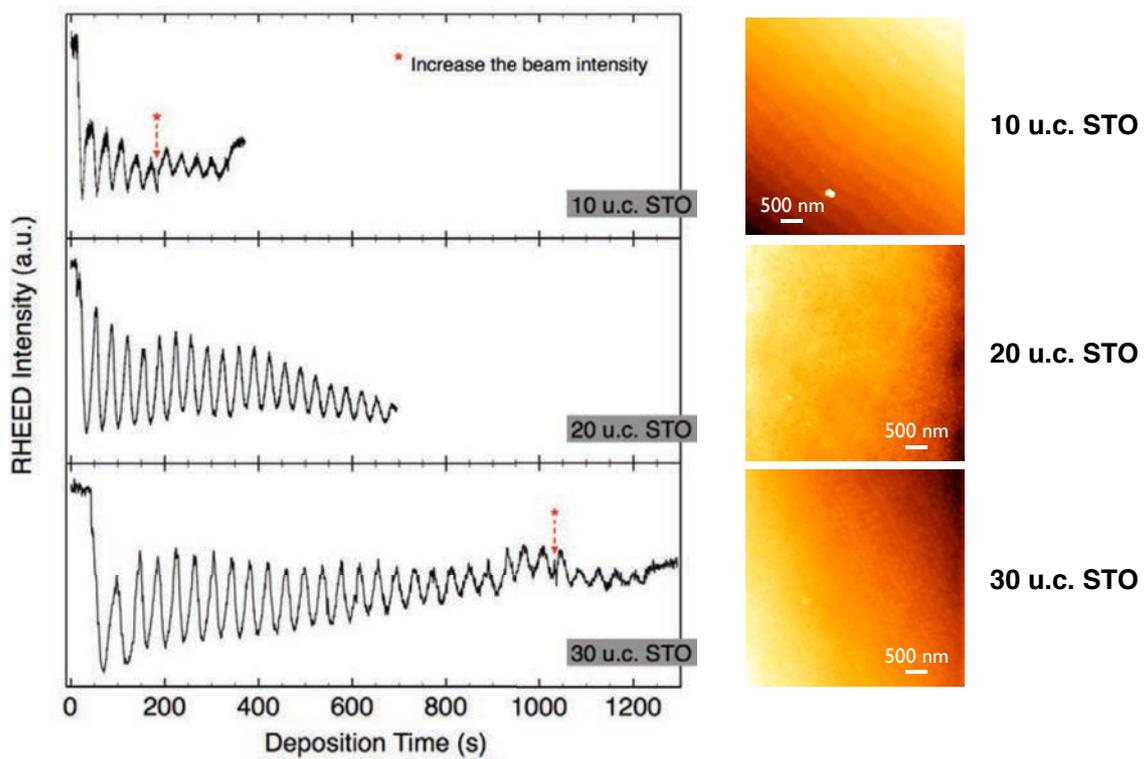

**Figure S1.** Growth characterization of STO films grown at 800°C: the RHEED intensity oscillations and the AFM topographic images.

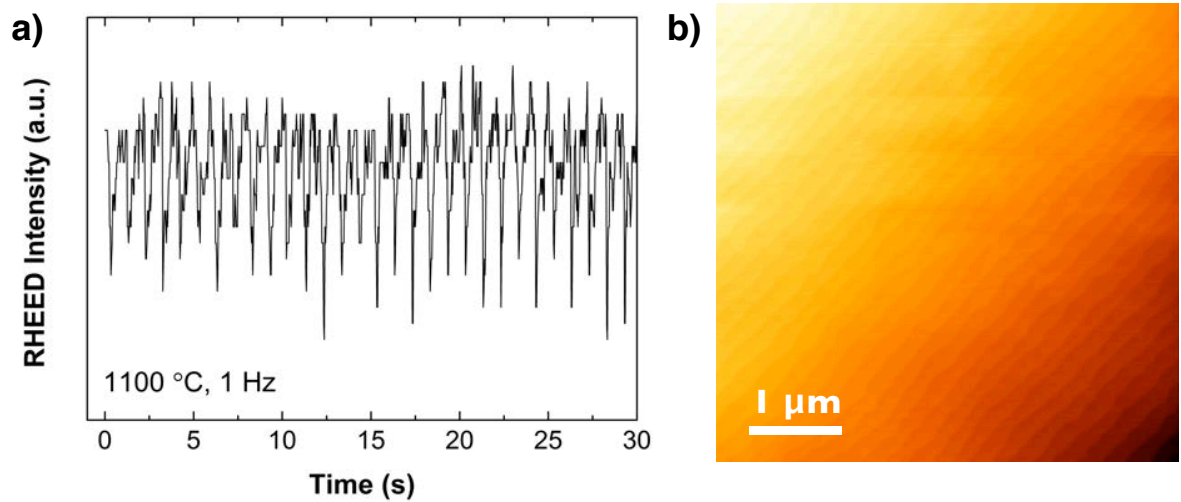

**Figure S2.** Growth characterization of STO films grown at 1100°C: the monitored RHEED intensity and the AFM topographic image.

**Transport characterization**

Detailed transport data for samples with STO layers grown at 800°C are shown in Figures S3 and S4: the sheet resistance and $-1/eR_H$ as a function of temperature are displayed for *in-situ* and *ex-situ* samples, respectively.

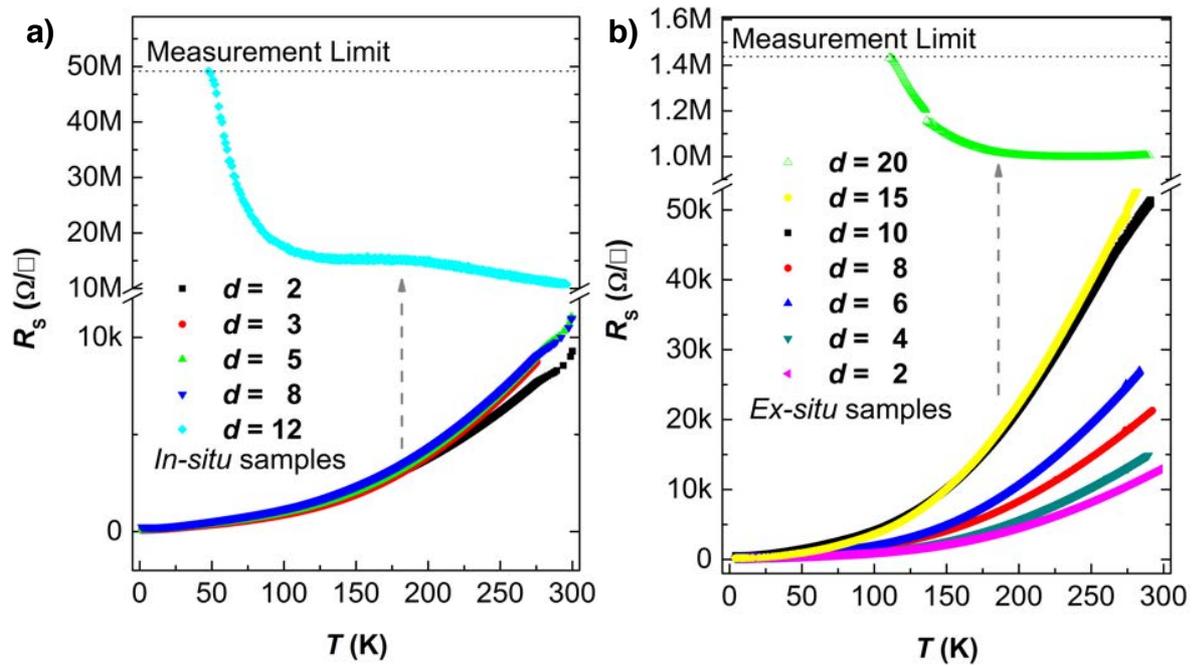

**Figure S3.** Sheet resistance as a function of temperature for *in-situ* samples and *ex-situ* samples with STO layers grown at 800°C.

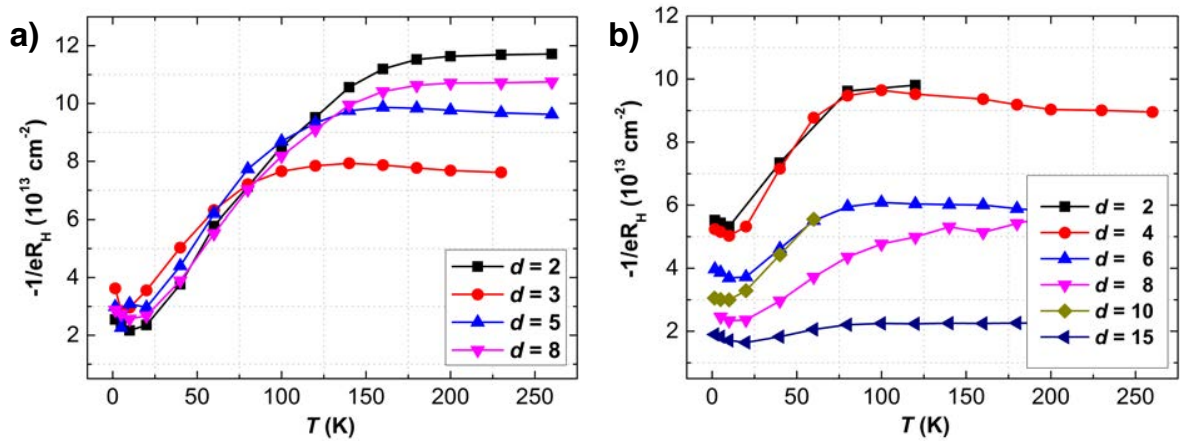

**Figure S4.** -1/e$R_H$ as a function of temperature for *in-situ* samples and *ex-situ* samples with STO layers grown at 800°C.

**Friction Force Microscopy Test**

In order to test the suitability of FFM as a tool for probing SrO segregation on the STO layer surface, we deposited a fraction of a SrO layer (a few laser pulses) on a STO substrate from a powder $SrO_x$ target as a test sample. A deposition temperature of 800°C and an oxygen pressure of $8\times10^{-5}$ Torr were used. Pits and small islands could be seen in the AFM topography after the deposition while no contrast could be resolved in FFM images. This may be due to the fact that the typical size of the SrO islands is below the FFM resolution. We then performed an annealing treatment (without BHF etching) using the conditions described in section *Experimental*. After the annealing, clear friction contrast can be seen in the FFM image (Figure S5). FFM images were obtained by subtracting the lateral retrace scan from the lateral trace scan. Contribution originating from the step edges can thus be diminished (see Figures S5c and S5d), leaving the signal purely related to the different friction coefficients (see Figures S5g and S5h).

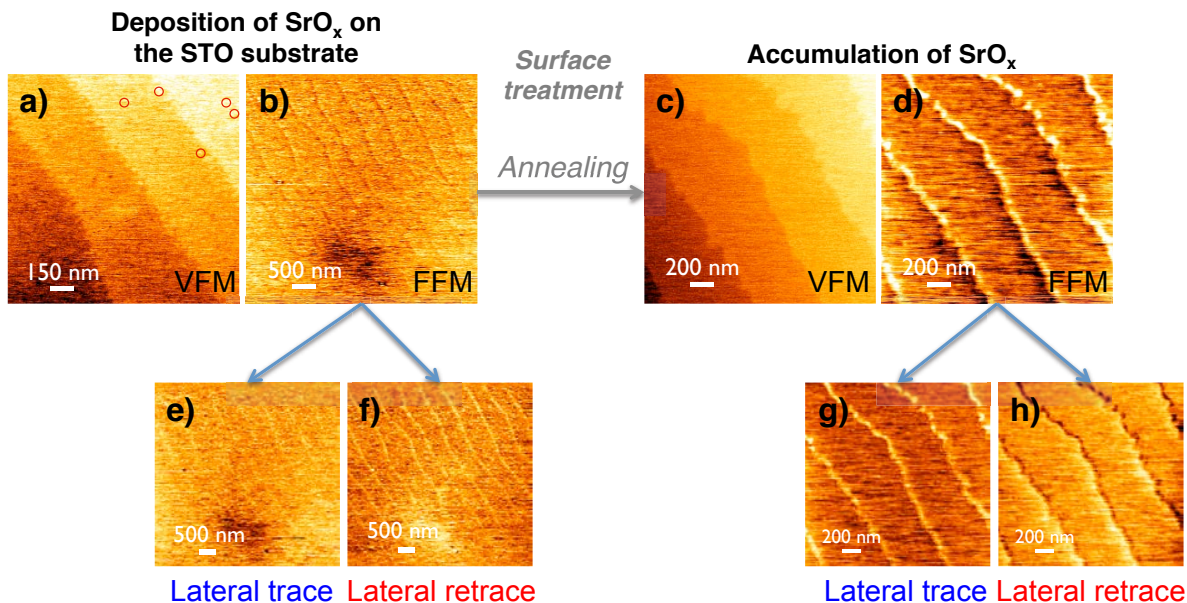

**Figure S5.** Surface characterization of $SrO_x$ species on STO substrates. a) topography (VFM) and b) chemical contrast (FFM) images of the STO surface after the deposition of a sub-monolayer of SrO. Red open circles indicate the presence of pits and small islands in the VFM image. No clear

contrast is observed in FFM images. c) VFM and d) FFM images of the sample after an annealing treatment. The contrast in the FFM image reveals the presence of SrO species on the surface accumulated at the step edges. e)-h) Lateral trace and retrace scans used to obtain the friction force images b) and d).